\documentclass[aps,prl,superscriptaddress,letterpaper,showpacs,amsmath,amssymb,reprint,floatfix,longbibliography]{revtex4-2}%
\usepackage{graphicx}
\usepackage{stmaryrd}
\usepackage{amsfonts}
\usepackage{siunitx}
\usepackage{xcolor}
\usepackage{physics} 
\usepackage[colorlinks=true,citecolor=blue,urlcolor=red]{hyperref}
\usepackage{orcidlink}

\begin{document}
\title{Hyperspectral electromechanical imaging at the nanoscale: Dynamical backaction, dissipation and quantum fluctuations}

\affiliation{CNRS, Inst. NEEL, "Nanophysique et semiconducteurs" group, 38000 Grenoble, France}
\affiliation{Université Paris-Saclay, CNRS, ENS Paris-Saclay, CentraleSupélec, LuMIn, 91405, Orsay, France}
\affiliation{Institut Universitaire de France, 1 rue Descartes, 75231 Paris, France}
\affiliation{Univ. Grenoble Alpes, CNRS, Grenoble INP, Institut N{\'e}el, F-38000 Grenoble, France}
\affiliation{Univ. Grenoble Alpes, CEA, IRIG, PHELIQS, “Nanophysique et semiconducteurs“ Group, F-38000 Grenoble, France}
\affiliation{Univ Lyon, ENS de Lyon, CNRS, Laboratoire de Physique, F-69342 Lyon, France}
\affiliation{School of Physics and Astronomy, University of Nottingham, Nottingham, NG7 2RD, United Kingdom}

\author{C. Chardin}
\affiliation{School of Physics and Astronomy, University of Nottingham, Nottingham, NG7 2RD, United Kingdom}

\author{S. Pairis}
\affiliation{Univ. Grenoble Alpes, CNRS, Grenoble INP, Institut N{\'e}el, F-38000 Grenoble, France}

\author{S. Douillet}
\affiliation{Univ. Grenoble Alpes, CNRS, Grenoble INP, Institut N{\'e}el, F-38000 Grenoble, France}

\author{M. Hocevar\,\orcidlink{0000-0001-5949-5480}} 
\affiliation{CNRS, Inst. NEEL, "Nanophysique et semiconducteurs" group, 38000 Grenoble, France}
\affiliation{Univ. Grenoble Alpes, CNRS, Grenoble INP, Institut N{\'e}el, F-38000 Grenoble, France}

\author{J. Claudon\,\orcidlink{0000-0003-3626-3630}}
\affiliation{Univ. Grenoble Alpes, CEA, INAC, PHELIQS, “Nanophysique et semiconducteurs“ Group, F-38000}

\author{J.-P. Poizat}
\affiliation{CNRS, Inst. NEEL, "Nanophysique et semiconducteurs" group, 38000 Grenoble, France}
\affiliation{Univ. Grenoble Alpes, CNRS, Grenoble INP, Institut N{\'e}el, F-38000 Grenoble, France}

\author{L. Bellon\,\orcidlink{0000-0002-2499-8106}}
\affiliation{Univ Lyon, ENS de Lyon, CNRS, Laboratoire de Physique, F-69342 Lyon, France}

\author{P. Verlot\,\orcidlink{0000-0002-5105-3319}}
\email{pierre.verlot@universite-paris-saclay.fr}
\affiliation{Université Paris-Saclay, CNRS, ENS Paris-Saclay, CentraleSupélec, LuMIn, 91405, Orsay,
France}
\affiliation{Institut Universitaire de France, 1 rue Descartes, 75231 Paris, France}

\begin{abstract}
We report a new scanning nanomechanical noise microscopy platform enabling to both heat and acquire the fluctuations of mechanical nanostructures with nanometric resolution. We use this platform to image the thermally activated nanomechanical dynamics of a model system consisting of a $40\,\mathrm{nm}$ diameter single-defect nanowire, while scanning a localized heat source across its surface. We develop a thermal backaction model, which we use to demonstrate a close connection between the structure of the nanowire, its thermal response, its dissipation and its fluctuations. We notably show that the defect behaves as a single fluctuation hub, whose e-beam excitation yields a far off-equilibrium vibrational state, largely dominated by the quantum fluctuations of the heating source. Our platform is of interest for future quantitative investigation of fundamental nanoscale dynamical phenomena, and appears as a new playground for investigating quantum thermodynamics in the strongly dissipative regime and at room temperature.
\end{abstract}

\date{\today}

\maketitle

\section*{Introduction}

Thermally activated nanomotion has a pivotal importance in nature, ranging from decoherence in quantum systems \cite{wilson2015measurement,riedinger2018remote} to controlling chemical reactions and elementary life processes \cite{wang2010motion,tu2017motion}: Owing to their very reduced dimensions, nanoscale systems generally display very high thermal resistances \cite{cahill2003nanoscale}, resulting in a strongly enhanced mechanical response to thermal phenomena, even at very low heating power \cite{zhang2013nanomechanical,tavernarakis2018optomechanics,blaikie2019fast}.
While this unique efficiency of nanoscale thermo-mechanical energy conversion and the corresponding central role of thermally induced nanomechanical dynamics being well established and acknowledged in nanophysics, the underlying mechanisms remain largely unknown, which essentially stems from the so far unsurpassed difficulty to image thermally activated internal dynamics with sufficient resolution. 

Here we overcome this stalemate and report the first experiment enabling to spatially image the thermally activated nanomechanical dynamics of nanomechanical systems. Using the electron beam (e-beam) of a scanning electron microscope (SEM) both for local heating and detection of the mechanical fluctuations, we are able to construct a nanometre-resolved hyperspectral of a $40\,\mathrm{nm}$ diameter Indium-Arsenide (InAs) nanowire (NW), that is an image where each pixel represents a spectrum of the NW's thermally-induced vibrational dynamics. We develop a general model, which enables to establish a close link between the structure of the NW, its thermal response, its dissipation and the observed nanomechanical fluctuations.

\begin{figure*}[t!]
\includegraphics[width=17cm]{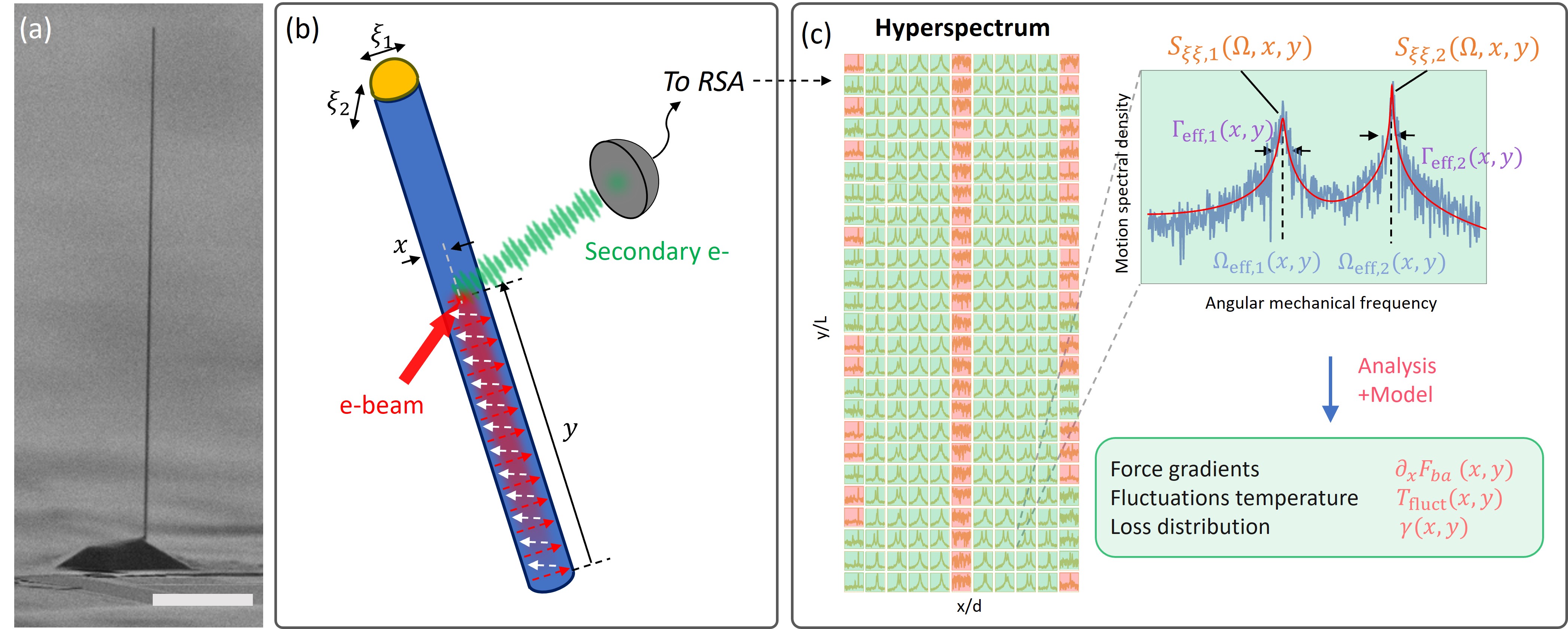} \centering 
\caption{\textbf{Sample layout and principle of the experiment.} (a) SEM micrograph showing a typical $40\,\mathrm{nm}$ diameter InAs nanowire similar to that used in this work.  Scale bar is $2\,\mu\mathrm{m}$. (b) Schematic depicting the principle of the experiment. A focused electron beam is swept accross the surface of the nanowire, with the horizontal and vertical coordinates $x$ and $y$ being scanned from left to right and bottom to top, respectively. Secondary electrons fluctuation spectra are recorded at each position using a real-time spectrum analyzer (RSA), which are used (c) to construct the hyperspectral image representing the main basis of this work. The spectra are processed so as to extract their characteristic features, including the effective mechanical angular frequencies $\Omega_{\mathrm{eff},i}$ and damping rates $\Gamma_{\mathrm{eff},i}$, and the calibrated spectral amplitudes $S_{\mathrm{\xi\xi},i}(\Omega,x,y)$ ($i\in\llbracket 1,2 \rrbracket$). Combined with our newly developed model, these quantities are used to determine the forces gradients, the motion temperature and loss distribution inside the nanowire.}
\label{Fig1}%
\end{figure*}

\section*{Hyperspectral thermomechanical imaging}

The NW investigated throughout this work has a length $L=7\,\mu\mathrm{m}$ and diameter $d=40\,\mathrm{nm}$, with corresponding tip effective mass $m\simeq 11.5\,\mathrm{fg}$ ($1\,\mathrm{fg}=10^{-18}\,\mathrm{kg}$) and fundamental mechanical resonance frequency $\simeq 455\,\mathrm{kHz}$, in close agreement with that expected from Euler-Bernoulli beam theory $\Omega_0\simeq 0.879 d/L^2\sqrt{E/\rho}\simeq2\pi\times472\,\mathrm{kHz}$, with $\rho=5670\,\mathrm{kg}\,\mathrm{m}^{-3}$ the mass density and $E=97\,\mathrm{GPa}$ the Young modulus of InAs~\cite{gandolfi2022ultrafast}. Details on the samples fabrication can be found in Method. The NW is mounted in a field emission SEM operating with an average probe current set to $\overline{I^{\mathrm{in}}}=300\,\mathrm{pA}$ and an acceleration voltage $V=3\,\mathrm{kV}$. The SEM simultaneously serves to heat the NW and detect its thermally induced vibration fluctuations \citep{Buks2000,Nigues2014a,pairis2019shot,cretu2022direct}. The main concept of the experiment is schematically depicted on Fig.~\ref{Fig1}(c): The electron beam is swept through the surface of the NW, from left to right and from bottom to top. Both DC and AC components  of the secondary electrons (SEs) current are recorded at position $(x,y)$ ($x$ and $y$ denoting the horizontal and vertical coordinates of the electron beam, see Fig.~\ref{Fig1}(c)), during the $\sim 10\,\mathrm{ms}$ e-beam dwell time, using an oscilloscope and a real-time spectrum analyzer (RSA), respectively). With these settings, the full NW scan requires $\sim 330\,\mathrm{s}$ of intermittent e-beam exposure time.

\begin{figure*}[t!]
\includegraphics[width=17cm]{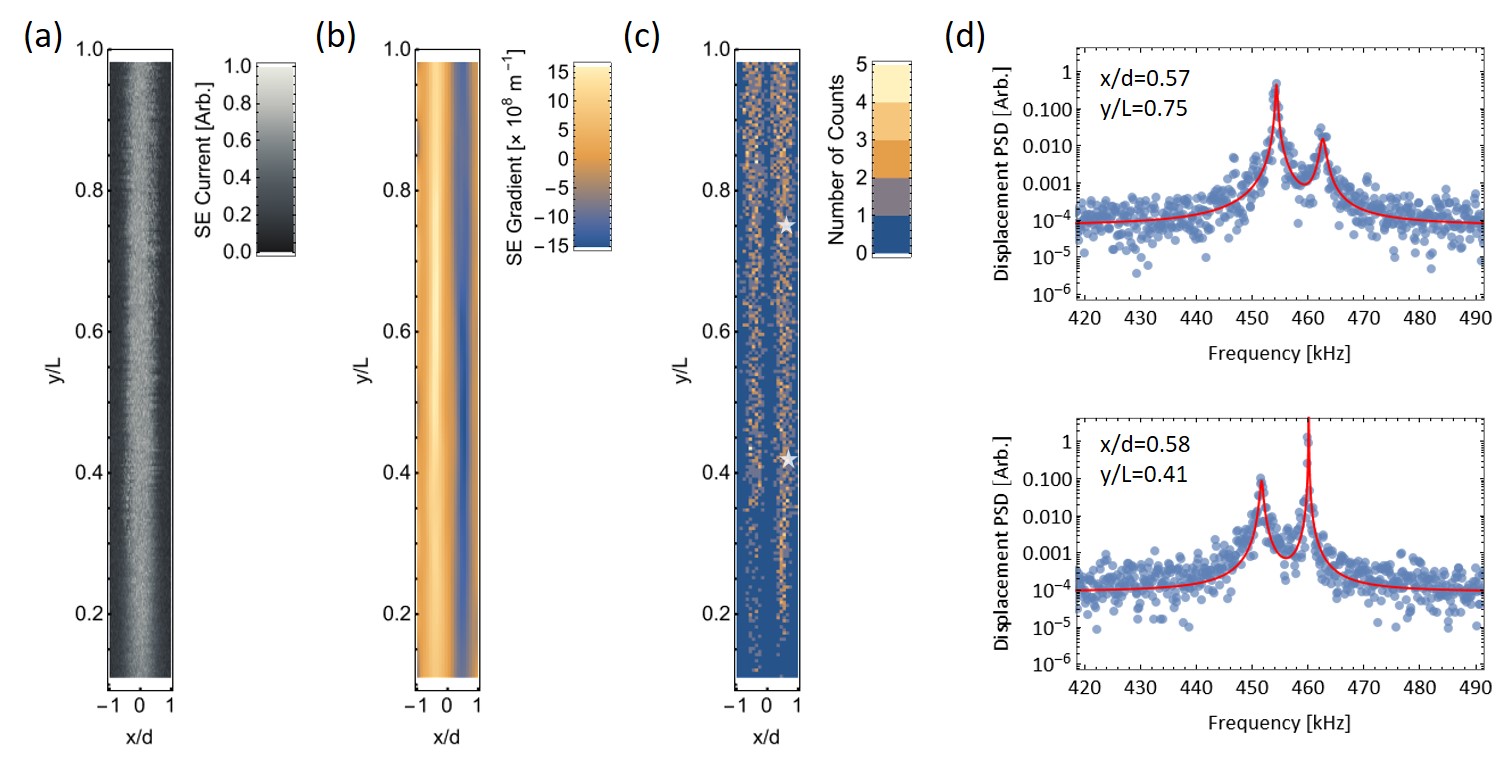} \centering
\caption{\textbf{Hyperspectral thermomechanical imaging with nanometric resolution.} (a) DC component of the secondary electrons current recorded accross the surface of a nanowire with length $L=7\,\mu\mathrm{m}$ and diameter $d=40\,\mathrm{nm}$. (b) Secondary electrons gradient with respect to the horizontal scanning direction $x$, proportional to the measurement sensitivity. (c) Number of contributing spectra as a function of the scanning position. The $74\,\mathrm{nm}\times 6.1\,\mu\mathrm{m}$ image is binned into $21\times201$ pixels, corresponding to a transverse resolution of $3.5\,\mathrm{nm}$. The star marks indicate the coordinates at which the spectra shown in Fig.~\ref{Fig2}(d) have been recorded. (d) Electromechanical spectra typically acquired to construct the electrothermal hyperspectral image. The double Lorenzian fits (solid lines) are used to extract the eigenfrequencies, effective dampings and temperatures associated with both vibrational directions $\xi_1$ and $\xi_2$ (see Fig.~\ref{Fig1}). }%
\label{Fig2}%
\end{figure*}

Figure \ref{Fig2}(a) shows the NW image reconstructed from the recorded SE DC current, displaying the large contrast characteristic of SEM imaging. Figure \ref{Fig2}(b) shows the associated SE gradient in the horizontal sweeping direction, which defines the nanomechanical motion sensitivity. Large gradients are found around the sidewalls of the NW $x/d=\pm1/2$, where the SE current quickly changes from zero to its maximum value with the horizontal coordinate. Conversely, the sensitivity cancels around the axis of the NW, where the SE current weakly depends on the e-beam position. Figure \ref{Fig2}(d) shows two typical mechanical spectra acquired on the right side of the NW, at the coordinates highlighted by the stars on Fig.~\ref{Fig2}(c). Two peaks are observed, corresponding to the non-degenerate fundamental flexural mode in each eigendirection of vibration  (see Fig.~\ref{Fig1}(c)), both having non-zero projection in the transverse measurement plan. 

Figure ~\ref{Fig2}(c) shows the number of electromechanical spectra successfully acquired across the surface of the NW. Due to the relatively limited dwell time, a spectrum is counted in as soon as both transduced mechanical peaks are simultaneously resolved with a signal-to-noise ratio $\geq 10\,\mathrm{dB}$, so as to limit the influence of the background noise. Here the frame was divided into $21\times201$ pixels in order to optimize both spectral and spatial resolutions of the electromechanical hyperspectrum, defined as the representation of the SE spectrum as a function of the e-beam coordinates. The number of counts increases towards the tip of the NW, where the motion amplitude tends to be larger than around the clamping region. In addition, the number of counts vanishes on the axis of the NW, where the motion sensitivity is very low. Remarkably, nonzero counts are continuously recorded across $85\%$ of the NW length, despite a $740$ fold increase of the effective mass from top to bottom. 

\begin{figure*}[t!]
\includegraphics[width=17cm]{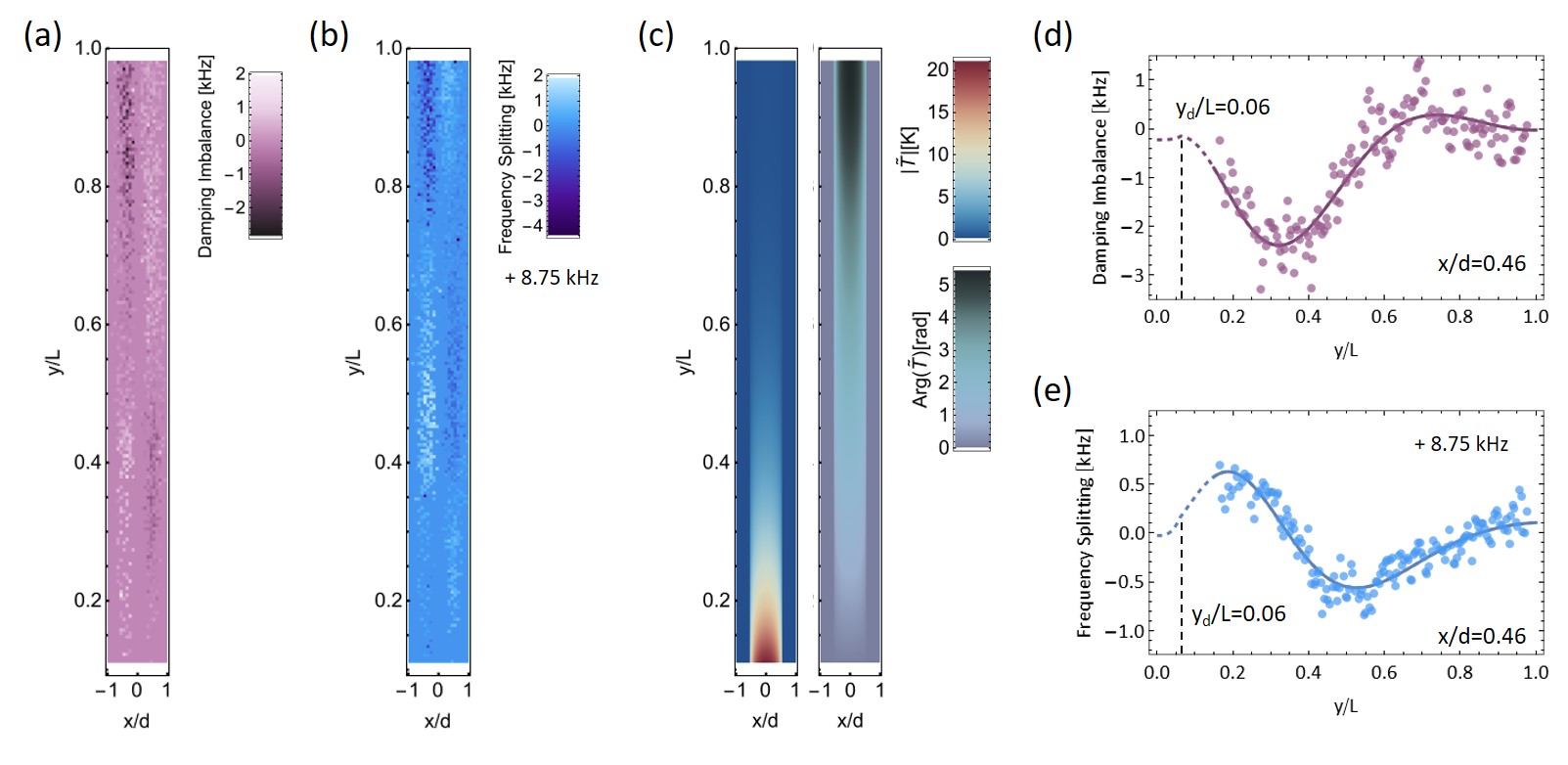} \centering
\caption{\textbf{Dynamical backaction from a single localized defect.} (a) Measured damping imbalance and (b) frequency splitting across the nanowire surface. (c) Amplitude (left) and phase (right) models of the temperature elevation $\tilde{T}$ at the defect position $y_{\mathrm{d}}/L=0.06$ (see Eq.~\ref{eq:4}) as a function of the e-beam position. (d-e) Longitudinal evolution of both the damping imbalance and frequency splitting, respectively ($x/d=0.46$). Dots are the experimental values inferred from fitting the recorded electromechanical spectra. Solid lines correspond to the single-defect electromechanical backaction model best fit, yielding to a defect position $y_{\mathrm{d}}/L=0.06$.}%
\label{Fig3}%
\end{figure*}

\section*{Dynamical backaction from a single defect}

We first investigate the spatial distribution of the thermal force gradients $\partial_xF_{\mathrm{ba}}$ and their dynamical effects on resonant mechanical motion. Force gradients act as additional restoring forces, with their in-phase and out-of-phase components $\mathrm{Re}[\partial_xF_{\mathrm{ba}}]$ and $\mathrm{Im}[\partial_xF_{\mathrm{ba}}]$ affecting the mechanical resonance frequency and damping rate, respectively. This phenomenon is also known as dynamical backaction \cite{aspelmeyer2014cavity} and is at the basis of atomic force microscopy~\cite{binnig1986atomic}.

A general caveat of frequency-based gradient microscopy is that off-resonant contributions, such as absorption/desorption induced mass changes, may sizeably affect the evolution of resonant properties. Here we circumvent this issue by taking advantage of the vectorial motion sensitivity of e-beam electromechanical coupling, which enables to reject the common-mode noise by considering the differential changes in the mechanical susceptibilities between the two vibrational directions (see Methods). Figures \ref{Fig3}(a) and \ref{Fig3}(b) show both the damping imbalance $\left\{\Gamma_{\mathrm{eff},2}(x,y)-\Gamma_{\mathrm{eff},1}(x,y)\right\}/2\pi\propto\mathrm{Im}[\partial_xF_{\mathrm{ba}}]$ and the centered frequency splitting $\left\{\Omega_{\mathrm{eff},2}(x,y)-\Omega_{\mathrm{eff},1}(x,y)-\Delta\Omega\right\}/2\pi\propto\mathrm{Re}[\partial_xF_{\mathrm{ba}}]$ of the mechanical resonances across the surface of the NW ($\llbracket 1,2 \rrbracket$ standing for the lower and higher frequency modes, respectively, and $\Delta\Omega$ the mean frequency splitting, see Methods). 

Two prominent features can be observed. Firstly, both the dissipative (Fig. \ref{Fig3} (a)) and dispersive (Fig. \ref{Fig3} (b)) contributions are antisymmetric with respect to the NW's axis. This is a distinctive signature of a \emph{unidirectional}, thermally-activated deformation mechanism, similar to those found in bimorph actuators~\cite{pedrak2003micromachined,Nigues2014a,tavernarakis2018optomechanics}: The temperature elevation yields to a non-zero, one-way bending of the NW via thermoelastic coupling, which is generally interpreted as the consequence of unavoidable structural inhomogeneities, whose relative weight is increased at the nanoscale.  The second, intriguing property shown on Figs. \ref{Fig3}(a-b) is a modulation of the thermal force gradients, in the longitudinal direction. This is in stark contrast with previous reports~\cite{Nigues2014a,tavernarakis2018optomechanics}, and appears to contradict the conventional approach to thermal backaction as being essentially set by the overlap between the temperature field and the strain distribution ~\cite{ramos2012optomechanics,primo2021accurate,ortiz2024influence}. In the present case of a cantilever fundamental flexural mode, such model would predict a thermal backaction force which monotonously increases with the longitudinal position of the heating source, which clearly departs from our experimental results.
  
To explain the observed behaviour, we develop an alternative, continuous thermal backaction model, the details of which can be found in \cite{Bellon2024}. With $(x,y)$ denoting the e-beam position, we assume that the effect of the induced frequency-dependent temperature elevation field $\widetilde{T}(x,y,y',\Omega)$ is to change the local curvature, resulting in a distributed force field given in Fourier space by:

\begin{eqnarray}
f_{\mathrm{ba}}(x,y,y',\Omega)&=&\partial_{y'}^2[E\mathcal{I}\beta(y',\Omega) \widetilde{T}(x,y,y',\Omega)],\label{eq:1}
\end{eqnarray}
with $\Omega$ the angular Fourier frequency, $E\mathcal{I}$ the flexural rigidity of the nanowire ($\mathcal{I}$ its second moment of area), and $\beta$ the thermoelastic response function of the NW ($y'$ the vertical coordinate). $\beta$ can be viewed as the NW's local curvature change per unit temperature \cite{Bellon2024}. The temperature elevation field is determined by solving the heat equation \cite{Bellon2024}, which reads in Fourier space as:
\begin{eqnarray}
\partial_{y'}^2\widetilde{T}=\alpha^2\widetilde{T},\label{eq:2}
\end{eqnarray}
with $\alpha^2=i\Omega\rho c/\kappa$, $\kappa$ the heat conductivity and $c$ the specific heat. Finally, the backaction force $F_{\mathrm{ba}}$ is given by the overlap between the distributed force field and the mechanical mode shape $\phi(y')$:
\begin{eqnarray}
F_{\mathrm{ba}}(x,y,\Omega)&=&\int_0^L\mathrm{d}y'f_{\mathrm{ba}}(x,y,y',\Omega)\phi(y').\label{eq:3}
\end{eqnarray}

The strong lateral confinement of the electromechanical interaction results in a large backaction force gradient $\partial_xF_{\mathrm{ba}}$, which significantly changes both the effective mechanical resonance frequency and the damping rate of the NW~\cite{metzger2004cavity,Arcizet2006,aspelmeyer2014cavity}. From Eqs. \ref{eq:1} and \ref{eq:3} , we see that those changes are expected to critically depend on the function $\beta$: Testing the model against the experimental data shown in Fig.~\ref{Fig3} therefore enables us to address the nature of the thermoelastic process, in principle.

We first considered the case of uniform thermal bending, that is $\beta(y')$ constant, independent of the longitudinal coordinate $y'$. This hypothesis is generally used for modelling the thermal deformations of bimorph structures~\cite{pedrak2003micromachined,ikuno2005thermally} and may be justified by the presence of a few nanometer-thick oxyde layer at the surface of the NW. However, this assumption definitely fails to describe the behaviour observed on Figure \ref{Fig3} \cite{Bellon2024}. We subsequently implement a radically different thermoelastic function, which we assume to correspond to a single defect located at a certain longitudinal position $y_{\mathrm{d}}$, $\beta(y',\Omega)=\beta_{\mathrm{d}}e^{i\Omega\tau_{\mathrm{d}}}\delta(y'-y_{\mathrm{d}})$, with $\beta_{\mathrm{d}}$ the thermal bending amplitude, $\delta$ the Dirac delta function, and with the complex exponential term accounting for the anelastic time response of the defect~\cite{choudhuri2007thermoelastic}. Under these conditions, the expression of the backaction force becomes:
\begin{eqnarray}
F_{\mathrm{ba}}(x,y,\Omega)&=&E\mathcal{I}\phi''(y_{\mathrm{d}})\beta_{\mathrm{d}}e^{i\Omega\tau_{\mathrm{d}}}\widetilde{T}(x,y,y_{\mathrm{d}},\Omega).\label{eq:4}
\end{eqnarray}

Eq.~\ref{eq:4} shows that the backaction force is simply proportional to the bending induced by the temperature elevation at the defect position (see Fig.~\ref{Fig3}(c)). We quantitatively try this model on longitudinal slices of the damping imbalance and frequency splitting ($x/d=0.46$), which are shown on Fig.~\ref{Fig3}(d-e). The solid lines are theoretical fits, respectively obtained from the imaginary and real parts of the lateral gradient of Eq.~\ref{eq:4} (see Methods). A remarkable agreement is found, with a defect location $y_{\mathrm{d}}\simeq0.06\,L$, that is $\simeq 420\,\mathrm{nm}$ above the clamping region, and an anelastic response time $\tau_{\mathrm{d}}\simeq68\,\mathrm{ns}$. This result can be further interpreted as follows: The position changes of the NW around the average position of the e-beam create a modulation of the heating power. This modulation feeds back to the mechanical motion, with a phase $\varphi_{\mathrm{ba}}(y)\sim(y-y_{\mathrm{d}})|\alpha(\Omega_0)|/\sqrt{2}-\Omega_0\tau_{\mathrm{d}}$, essentially set by the heat diffusion length $\sim 1/|\alpha[\Omega_0]|\simeq 0.8\,\mu\mathrm{m}$ (Eq.~\ref{eq:2}).

We notably see that around $y\simeq 0.35L$, the backaction phase is $\varphi_{\mathrm{ba}}\simeq\pi/2$ (see Fig.~\ref{Fig3}(c)), corresponding to a purely dissipative feedback force~\cite{gavartin2012hybrid}, with no associated frequency change. Conversely, $\varphi_{\mathrm{ba}}(y\simeq0.55L)\simeq\pi$, a situation for which the frequency variation is maximal, while the dissipative contribution is zero. Furthermore, as the e-beam position is swept towards the tip of the NW, the effect of the thermal wave becomes increasingly damped around the defect location (see Fig.~\ref{Fig3}(c)), $\propto \exp\left[{-|\alpha(\Omega_0)|(y-y_{\mathrm{d}})/\sqrt{2}}\right]$, which explains the vanishing of the backaction-induced damping and frequency changes.

\section*{Quantum-driven fluctuations and dissipation} 

\begin{figure*}[t!]
\includegraphics[width=17.8cm]{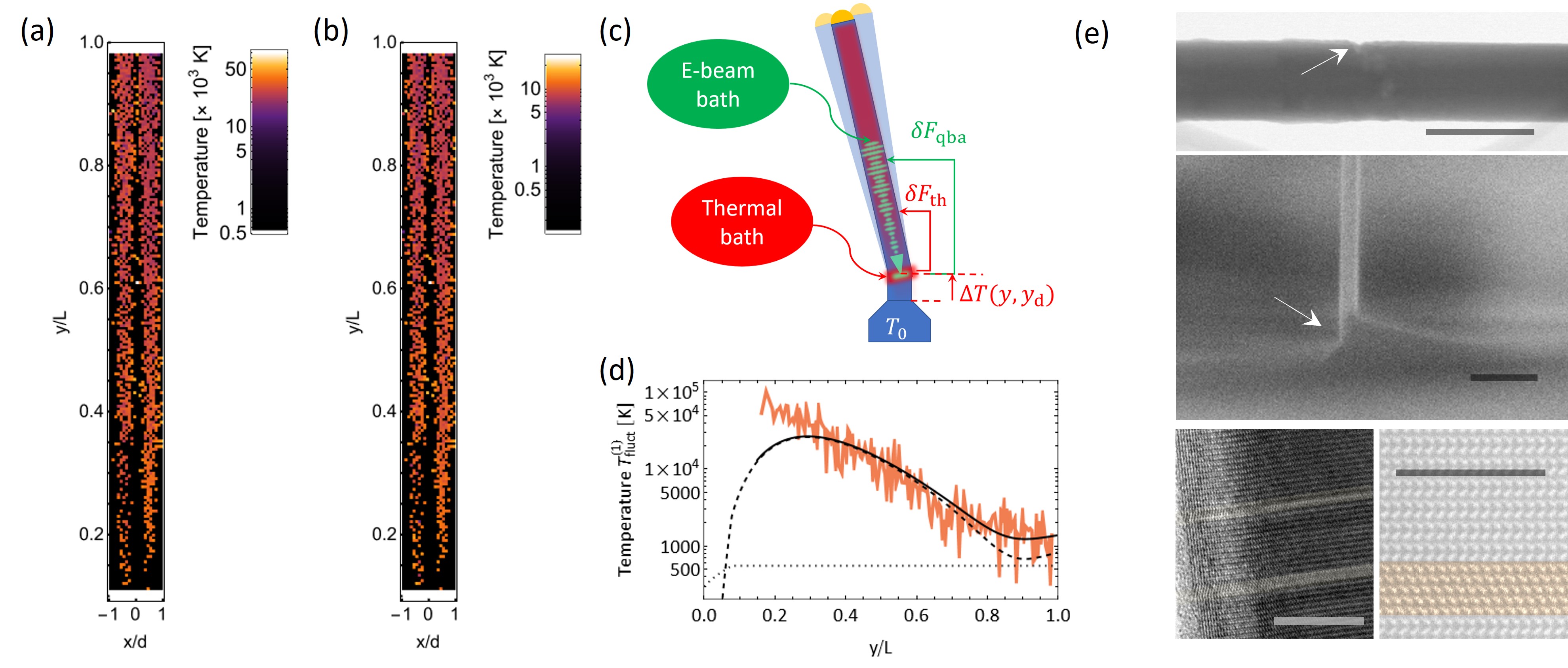} \centering
\caption{\textbf{Motion temperatures and quantum effects.} (a-b) Calibrated motion temperatures $T_{\mathrm{fluct}}^{(1)}$ and $T_{\mathrm{fluct}}^{(2)}$ associated with mode 1 (a) and 2 (b), respectively. (c) A localized thermoelastic defect as a single fluctuation hub. The defect essentially defines the dissipation of the whole mechanical structure, therby acting as a single thermomechanical fluctuations emmitter. Concurrently, the quantum current fluctuations of the e-beam are thermally transduced down to the defect, which converts them into motion via the thermoelastic mechanism.  (d) Longitudinal evolution of the motion temperature associated with mode 1 ($x/d=0.46$). The dashed line is the amplitude-fitted quantum backaction model, with $y_{\mathrm{d}}/L=0.08$. The dotted line represents the NESS temperature model ($y_{\mathrm{d}}/L=0.08$). The solid line is the sum of both. (e) Three types of defects encountered with our NWs. Top: Transmission Electron Micrograph showing an indentation defect. Middle: Scanning Electron Micrograph showing a clamping defect, with assymmetric boundaries between the NW and its pyramidal basis being apparent. Bottom Left: Faceting defects, appearing as local diameter variation of the NW (brighter regions). Such defects are typically associated with the insertion of the cubic Zinc Blende structure within the regular hexagonal Wurtzite structure, as shown on the figure opposite. Bottom Right: High Resolution Transmission Electron Micrograph showing a Zinc-Blende insertion in the Wurtzite lattice (see the highlighted, orange stripe), associated with a crystalline discontinuity. Scale-bars are 100 nm, 300 nm, 10 nm (bottom left) and 5 nm, respectively.}
\label{Fig4}
\end{figure*}

The identification of a thermoelastic defect around the basis of the NW suggests to consider its contribution to the mechanical dissipation, and therefore to its impact on the thermal fluctuations. In particular, the notion that the losses be concentrated in a single, known location raises the opportunity to test the fluctuation-dissipation relation at the microscopic scale, and to spatially pinpoint the origin of the thermal fluctuations, which may open new access towards thermal bath control.

Figures~\ref{Fig4}(a-b) show the temperature $T_{\mathrm{fluct}}^{(1,2)}$ associated with the motion variance in each direction of motion (see Methods). The longitudinal evolution of $T_{\mathrm{fluct}}^{(1)}$ across the transverse beam position $x/d=0.46$ is magnified on Fig.~\ref{Fig4}(d). A very surprising, strong increase of the fluctuation temperature, up to unphysical material temperatures, is observed as the e-beam approaches the NW's clamping region. However, addressing the thermal state of a mechanical system subject to strong thermal gradients is a complex task, requiring advanced statistical physics tools enabling to generalize the fluctuation-dissipation theorem, that may sometimes yield to counter intuitive outcomes.

\paragraph{Thermal Non-Equilibrium Steady State.}
In particular, it has recently been shown that thermally driven fluctuations may acutely depend on both the loss mechanism and distribution, leading to thermal non-equilibrium steady states (NESS), whose temperature does not match that of the material~\cite{geitner2017low}. One important take-away from the NESS approach is that, in general, the motion temperature corresponds to the barycenter of the temperature elevation $\Delta T(y,y')$, weighted by the normalized, local dissipation rate $\gamma(y')$, $T_{\mathrm{NESS}}(y)=T_0+\int_0^L\mathrm{d}y'\Delta T(y,y')\gamma(y')$ (with $T_0=300\,\mathrm{K}$ the ambient temperature). For a point-like, thermoelastic loss process located at position $y_{\mathrm{d}}$, $\gamma(y')=\delta(y'-y_{\mathrm{d}})$. Consequently, the thermal NESS theory predicts that the associated thermal fluctuations are set by the temperature of the defect, $T_{\mathrm{NESS}}(y)=T_0+\Delta T(y,y_{\mathrm{d}})$. In other words, the defect acts as a single fluctuation port, powering the whole thermal motion of the NW. Moreover, the continuity of the thermal flux implies that heating the defect from above is not sensitive to the e-beam longitudinal position, that is $\Delta T(y\geq y_{\mathrm{d}},y_{\mathrm{d}})$ is constant \cite{Bellon2024}. Therefore for a  NESS, a constant fluctuation temperature $T_\mathrm{NESS}$ is expected to be measured over the full e-beam positions range accessible to our experiment ($y \gtrsim 0.11\,L>y_{\mathrm{d}}$).

Figure~\ref{Fig4}(d) shows that, while the motion temperature around the tip of the NW approximates that of a single-defect thermal NESS (dotted line, no fitting parameter being used), it rapidly diverges from the thermal NESS model as the e-beam getting closer to the NW's basis, which is confirmed all across the surface and for the second direction of motion (Figs.~\ref{Fig4}(a-b)): This demonstrates that the measured motion fluctuations are of a non-thermal nature.

\paragraph{Quantum Non-Equilibrium Steady State.}
The other main source of noise in our experiment is the electron beam itself, obtained via a field emission process, which exhibits Poissonian-type quantum fluctuations (shot noise) ~\cite{kodama2000shot}, proportional to the square root of the mean output electron flux $\overline{I^{\mathrm{in}}}/e$ ($e$ the positron's charge). The backaction force, proportional to the incident electron flux via the elevation temperature field therefore has some noise (also known as quantum backaction noise) \cite{Bellon2024}, which acts as an external, quantum fluctuation bath, with a force noise spectral density given by:
\begin{eqnarray}
S_{\mathrm{FF,q}}^{\mathrm{ba}}(x,y,\Omega)&=\frac{e}{\overline{I^{\mathrm{in}}}}\times|F_{\mathrm{ba}}(x,y,\Omega)|^2&.\label{eq:5}
\end{eqnarray}

The dashed line in Fig.~\ref{Fig4}(d) shows the quantum backaction temperature associated with the force noise spectral density set by Eq.~\ref{eq:5}, with the amplitude being the only fitting parameter. The solid line shows the compound sum of the quantum backaction and thermal NESS contributions, displaying very good agreement with the data.

Interestingly, these results reinforce the interpretation of the defect as a single noise port. The present experiment can indeed be viewed as displacing the fluctuation bath over the surface of the NW (see Fig.~\ref{Fig4}(c)). The quantum fluctuations of the e-beam are transmitted to the defect via an electrothermal conversion process, whose dynamics is essentially set by the heat equation (Eq.~\ref{eq:2}). Thereby, the heat wave generated around the apex of the NW is considerably damped when reaching the clamping region, resulting in a strong filtering of the e-beam fluctuations transmitted to the defect. The subsequent thermoelastic transduction of these quantum fluctuations produces weak displacements (dashed line in Fig.~\ref{Fig4}), the motion fluctuations corresponding essentially to a thermal NESS, with fluctuation temperature set by the defect, $\simeq T_0+\Delta T(y,y_{\mathrm{d}})$ (dotted line in Fig.~\ref{Fig4}). Conversely, the transmission of the e-beam fluctuations becomes optimal around the defect location, where they are directly feeding in, in real-time. The associated thermoelastic displacements are large, corresponding to high fluctuation temperatures $T_\mathrm{fluct}$.

\section*{Discussion}

\paragraph{Quantum effects in a hot dissipative mechanical system}
Fig.~\ref{Fig4}(d) shows that the quantum backaction drives the mechanical motion to temperatures exceeding $10\,000\,\mathrm{K}$, that is more than $20$ times the thermal noise level. This result may appear surprising, or even counterintuitive: so far, only a handful of optomechanical experiments have been able to demonstrate sizeable quantum backaction effects at room temperature, however to levels still comparable to that of the thermal noise~\cite{cripe2019measurement,magrini2022squeezed,militaru2022ponderomotive}. These reports have in common a weak mechanical dissipation, so as to minimize the thermal force noise, and the use of high probe powers ($\geq 100\,\mathrm{mW}$), corresponding to large photon fluxes, in excess of $10^{18}\,\mathrm{s}^{-1}$. Conversely, our experiment operates with a rather low quality factor $Q\sim 150$, a probe power $\sim 1\,\mu\mathrm{W}$ and an electron flux $\overline{I^{\mathrm{in}}}/e\simeq 1.9\times 10^{9}\,\mathrm{s}^{-1}$. This spectacular difference essentially explains because of two aspects. Firstly, the electrothermal backaction is parametrized by the thermal bending amplitude, which magnifies the effects of the power fluctuations on the mechanical motion. A better grasp of this parameter can be accessed by estimating the tip's static displacement associated with the high quantum backaction temperature, $\xi_{\mathrm{ba}}(L)=\frac{1}{\Omega_0^2}\times\sqrt{\frac{2\Gamma_0 k_B T_{\mathrm{ba}}\overline{I^{\mathrm{in}}}}{me}}\simeq115\,\mathrm{nm}$ for $T_{\mathrm{ba}}=10\,000\,\mathrm{K}$ ($\Gamma_0=\Omega_0/Q$ the intrinsic mechanical damping rate and $k_B$ the Boltzmann constant), that is around $1-2\%$ of the NW's length and is routinely observed in SEM with similar structures. The second aspect is owed to the much higher single particle energy ($\sim 10^3\,\mathrm{eV}$ vs $\sim 1\,\mathrm{eV}$ for electrons and photons, respectively), which enables to operate with reduced electron fluxes for a given incoming power. This reinforces the relative weight of quantum fluctuations ($\propto\sqrt{\overline{I^{\mathrm{in}}}}$) compared to mean fields ($\propto\overline{I^{\mathrm{in}}}$). In our case, the shot noise-induced power modulation depth experienced by the NW over its mechanical bandwith is $\sim\sqrt{e\Gamma_0/2\pi\overline{I^{\mathrm{in}}}}\simeq0.1\%$ (versus e.g. $\lesssim 10^{-7}$ for a $100\,\mathrm{mW}$ infrared laser beam).

\paragraph{Thermal vs Quantum noise}

As outlined above, the thermal and quantum backaction forces proceed from clearly distinct mechanisms. In the former case, the average thermal energy stored in the nanowire excites the random motion of the crystal matrix constituents. Because of the dissipation, a transfer of energy between the atoms localised at the defect's position and the NW acoustic modes sets in, yielding to Brownian motion. Conversely, the quantum backaction results from temperature fluctuations, proportional to the e-beam shot noise. The atoms localized at the defect position coherently respond to the resulting thermal waves, yielding to a local strain that propagates to the whole structure. Interestingly, the force spectral densities associated with both effects show the same dependency with respect to the input electron flux, proportional to it. This, however, stems from radically different grounds, respectively reflecting the incoherent nature of the thermalization process, and the quantum statistics of the electron beam. This reminds that, contrary to a widely spread idea, the power scaling of a measurement-driven dynamical phenomenon does not allow to conclude as to its quantum origin, in principle. Besides the spatial degree of freedom hereby exploited, e-beam electromechanical coupling offers a way to decouple thermal and quantum effects that is not available to the optical domain, that is to change the e-beam noise statistics while keeping the input power constant, by adapting the acceleration voltage. Thereby, operating with a low acceleration voltage and a high e-beam flux would enable to $\emph{decrease}$ the quantum backaction to the benefit of thermal effects, and the other way around, which results from the relatively increased weight of quantum fluctuations in the lower particle number limit, as discussed in the previous section. 

\paragraph{Defect morphology and dissipation}

Interestingly, the defect location determined from our analysis, in a region of large mechanical stress, is consistent with a degraded mechanical quality factor of around $Q\sim150$, more than an order of magnitude below the highest values measured from the same batch of NWs. This confirms the close relation between the dissipation, the (dissipative) thermoelastic backaction and the corresponding fluctuations of the NW. We also note that fitting the longitudinal evolution of the backaction gradients and quantum fluctuations to our model yield to distinct values for the position of the defect, of $y_{\mathrm{d}}\simeq 420\,\mathrm{nm}$ and $\simeq560\,\mathrm{nm}$ above the NW clamp, respectively. This essentially sets the accuracy of the defect location, on the order of a few times the diameter value. This result is somewhat unsurprising, as this reasonably corresponds to the footprints of both the thermal and strain boundaries.

While our model does not enable to determine the exact nature of the defect, it is however possible to identify several candidates, associated with very localized asymmetric strain distributions.  Figure 4(e) shows such defects typically observed with our fabrication process. The top image shows a scanning transmission electron microscope (TEM) micrograph of an InAs nanowire, exhibiting an indentation defect. The middle image is a SEM micrograph showing the asymmetric clamping of a NW: The InAs nanowire growth is generally accompanied by the formation of a pyramid at its basis due to adatom diffusion effects. The top facets of the pyramid grows at different rates, resulting in an asymmetric overlay of the nanowire. The bottom left image is a TEM micrograph showing faceting defects (brighter regions), which result from axial stacking faults depending on the growth conditions. Such stacking defects are magnified on the high resolution TEM micrograph of a NW crystalline structure, shown at the bottom right, and where small insertions of the cubic Zinc Blende structure inside the regular hexagonal Wurtzite structure can clearly be observed (highlighted, orange stripe). 

Note that, other asymmetric defects may be encountered, such as the deposition of a rough layer of native oxide surrounding the nanowire, or crystalline disorder resulting from recrystallization of the material following high energy electron beam imaging. These structures however essentially extend over surfaces/volumes comparable to that of the NW, as opposed to the very localized character of the defects shown on Fig. 4(e).           

\section*{Conclusion}
We have reported a new hyperspectral imaging experiment, based on a pixel-by-pixel analysis of the secondary electron fluctuations in a scanning electron microscope. We demonstrate that the ability of electron beams to create strong thermal gradients in a nanomechanical device gives rise to dynamical backaction, which can be used as a fingerprint of both the nature and spatial distribution of the loss mechanism. We exploit these new capabilities and address the origin of the dissipation in a $40\,\mathrm{nm}$ diameter, $7\,\mu\mathrm{m}$ long InAs semiconducting nanomechanical wire, which we identify as originating from a single defect localized around the NW clamping region. Based on the analysis of the spatial dependency of the nanomechanical motion temperature, we show that this defect behaves as the main fluctuation hub for the NW, both emmitting thermal energy and transmitting quantum fluctuations to the acoustic degrees of freedom. We notably observe that around the defect location, the transduction of the e-beam quantum fluctuations is so efficient as to generate a quantum backaction displacement noise more than $20$ times larger than the thermal motion. Our plateform appears as a new playground for quantum thermodynamics in the strongly dissipative regime and at room temperature. On a more technological level, our results may be exploited as a new tool for characterizing nanomaterials, and stresses the so far overlooked importance of quantum fluctuations in e-beam assisted fabrication processes~\cite{van2008critical,de2016review,gruber2019mass}.

\section{Acknowledgements}
We are grateful to Olivier Bourgeois, Eddy Collin, Alex Fontana and Andrew Armour for discussions, and Fabrice Donatini for help, assistance and advice at the early stage of this project. We thank Prof. Susheng Tan from the University of Pittsburgh for providing transmission electron microscope images of our InAs nanowires.
This work has been supported by the projects "QDOT" (ANR-16-CE09-0010) and "iMAGIQUE" 
(ANR-22-CE42-0022) from the French Agence Nationale de la Recherche, and by the project "Q-ROOT" 758794 from the European Research Council.

\section{Methods}
\paragraph{Samples fabrication}
Our NWs are grown perpendicularly on a (111)B InAs substrate by the vapour solid liquid method, using  diluted $20$ nm-gold  colloids (BBI Solution) as catalysts, in a molecular beam epitaxy setup. The growth parameters were set to an equivalent 2D growth rate of 0.2 nm/s, a V/III beam equivalent pressure ratio of 100 and a substrate temperature of $420^{\circ}\mathrm{C}$. This process results in a forest of nanowires (with a typical density of $10^4\,\mathrm{mm}^{-2}$), with each nanowire being surmounted by its own hemispherical gold catalyst droplet 
and featuring a very pure wurtzite crystal structure. 
\paragraph{Calibration of nanomechanical motion}
The electromechanical coupling is essentially sensitive to nanomechanical motion in the transverse plane, where the secondary electrons gradient is much larger. The resulting SEs fluctuations are given by:
\begin{eqnarray}
\delta I&=&\left(\frac{\mathrm{d}I}{\mathrm{d}x}\right)_{(x,y)}\xi+\delta I_{\mathrm{q}},\label{eq:M1}
\end{eqnarray}
with $\xi=\cos{\theta}\xi_1+\sin\theta\xi_2$ the projection of the two-dimensional nanomechanical motion $(\xi_1,\xi_2)$ in the transverse plane, $\mathrm{d}I/\mathrm{d}x$ the secondary electrons transverse gradient at position $(x,y)$ and $\delta I_{\mathrm{q}}$ the secondary electrons shot noise fluctuations. The same calibration procedure applies to both motion polarizations, and is further detailed below for $\xi_1$ (dropping the subscript $_1$ for ease of reading).

For angular Fourier frequency $\Omega$ around the first mechanical resonance $\Omega_1$, the spectrum of the secondary electrons current fluctuations can be approximated as:
\begin{eqnarray}
S_{\mathrm{II}}[\Omega]&\simeq&\left(\frac{\mathrm{d}I}{\mathrm{d}x}\right)^2_{(x,y)}\times \cos^2\theta\, S_{\mathrm{\xi\xi}}[\Omega]+S_{\mathrm{II,q}},\label{eq:M2}
\end{eqnarray}
with $S_{\mathrm{\xi\xi}}$ the displacement noise spectral density and $S_{\mathrm{II,q}}$ the SE current shot noise flat spectral density. For $x/d\simeq 0.46$ (corresponding to the transverse position at which the data shown on Figs. \ref{Fig3}-\ref{Fig4} have been acquired), the SEs current slope is close to maximum, $\mathrm{d}I/\mathrm{d}x\simeq2\eta\overline{I^{\mathrm{in}}}/d$ ($\eta\simeq0.3$ the secondary electron yield of InAs at $3\,\mathrm{keV}$~\cite{hussain2020monte}), and the average SEs current is about half of its maximum value, yielding to $S_{\mathrm{II,q}}\simeq\eta I^{\mathrm{in}}/2$. This direct relation between the shot noise level and the average SEs current enables to calibrate the motion spectrum in a self-contained manner, by comparing the resonant fluctuations to the off-resonant background:
\begin{eqnarray}
S_{\mathrm{\xi\xi}}[\Omega]&\simeq&\frac{1}{\cos^2\theta}\times \left(\frac{d}{4S_{\mathrm{II,q}}}\right)^2\times S_{\mathrm{II}}[\Omega].\label{eq:M3}
\end{eqnarray} 

Experimentally, we have access to an amplified transimpedance voltage $\delta \tilde{v}=G\delta I$, with the gain parameter $G$ determined from the off-resonant shot noise transduction, $G^2=2 S_{\mathrm{\tilde{v}\tilde{v},q}}/\eta\overline{I^{\mathrm{in}}}$. This leads to the expression we use for calibrating our spectra:
\begin{eqnarray}
S_{\mathrm{\xi\xi}}[\Omega]&\simeq&\frac{1}{2\eta\overline{I^{\mathrm{in}}}\cos^2\theta}\left(\frac{d}{2}\right)^2\times\frac{S_{\mathrm{\tilde{v}\tilde{v}}}[\Omega]}{S_{\mathrm{\tilde{v}\tilde{v},q}}}.\label{eq:M4}
\end{eqnarray}

The angle $\theta$ is determined by comparing the resonant values of the current spectral densities of both modes, $\cos^2\theta=\frac{1}{1+
\langle S_{\mathrm{II}}[\Omega_2]/S_{\mathrm{II}}[\Omega_1]\rangle \cot^2\theta_{\mathrm{ba}}}$, where $\theta_{\mathrm{ba}}$ denotes the angle between the backaction force and the first axis of vibration, see the Supplementary Note: "Determination of the backaction force orientation" ($\langle ... \rangle$ standing for averaging over the NW surface). 

\paragraph{Determination of the fluctuation temperature}

Around resonance ($\Omega\simeq\Omega_1$), the motion spectrum takes the following form:
\begin{eqnarray}
S_{\mathrm{\xi\xi}}[\Omega]&=&|\chi_{(x,y)}[\Omega]|^2S_{\mathrm{FF},1}[\Omega],\label{eq:M5a}\\
\chi_{(x,y)}[\Omega]&=&\frac{1}{m_{\mathrm{eff}}(y)\{\Omega_{\mathrm{eff},1}^2(x,y)-\Omega^2-i\Gamma_{\mathrm{eff},1}(x,y)\Omega\}},\nonumber\\
\label{eq:M5b}
\end{eqnarray}
with $S_{\mathrm{FF},1}$ the total force noise spectral density experienced by mode $1$, $\chi_{(x,y)}$ the position dependent nanomechanical suceptibility ($\Omega_{\mathrm{eff},1}$ and $\Gamma_{\mathrm{eff},1}$ the effective, dynamical backaction-modified mechanical resonance frequency and damping rate), and with $m_{\mathrm{eff}}(y)=\frac{\phi^2(L)}{\phi^2(y)}m$ the local effective mass. The fluctuation temperature is defined as:
\begin{eqnarray}
T_{\mathrm{fluct}}(y)&=&\frac{S_{\mathrm{FF},1}}{2m_{\mathrm{eff}}(y)\Gamma_{\mathrm{m}}k_B},\label{eq:M6}
\end{eqnarray} 
with $\Gamma_{\mathrm{m}}$ the intrinsic mechanical damping rate. The total force noise spectral density can be written as a function of the resonant value of the motion spectrum, $S_{\mathrm{FF},1}=m_{\mathrm{eff}}^2(y)\Gamma_1^2\Omega_1^2S_{\mathrm{\xi\xi}}[\Omega_1]$, from which we obtain the expression used for evaluating the fluctuation temperature displayed in the manuscript in Fig.~\ref{Fig4}(d):
\begin{eqnarray}
T_{\mathrm{fluct}}(y)&=&\frac{m\Gamma_1^2\Omega_1^2S_{\mathrm{\xi\xi}}[\Omega_1]}{2\Gamma_{\mathrm{m}}k_B}\frac{\phi^2(L)}{\phi^2(y)}
\end{eqnarray}

\paragraph{Common mode noise cancellation}
The NW undergoes alterations while being scanned, including e-beam induced absorption/desorption, to which the mechanical susceptibility is very sensitive because of the very low, femtogram mass. These alterations equally affect both vibration modes, such that it is possible to circumvent them by using common mode noise cancellation. This amounts to study the frequency splitting, and therefore the damping imbalance for maintaining the correct proportions between the dissipative and dispersive effects of the dynamical backaction:
\begin{eqnarray}
\Omega_{\mathrm{eff},2}-\Omega_{\mathrm{eff},1}&\simeq&\Delta\Omega-\frac{\mathrm{Re}\left(\partial_x F_{\mathrm{ba}}\right)}{2m_{\mathrm{eff}}\Omega_0}\left[\sin{\theta_{\mathrm{ba}}}-\cos{\theta_{\mathrm{ba}}}\right],\nonumber\\
\label{eq:M7a}\\
\Gamma_{\mathrm{eff},2}-\Gamma_{\mathrm{eff},1}&\simeq&\frac{\mathrm{Im}\left(\partial_x F_{\mathrm{ba}}\right)}{m_{\mathrm{eff}}\Omega_0}\left[\sin{\theta_{\mathrm{ba}}}-\cos{\theta_{\mathrm{ba}}}\right],\label{eq:M7b}
\end{eqnarray}
with $\Delta\Omega$ the mean frequency splitting. The damping rates being insensitive to the measurement-induced alterations, they can be exploited for determining the direction of the backaction force, yielding to $\theta_{\mathrm{ba}}\simeq -0.6\,\mathrm{rad}$ (see Supplementary Note: "Determination of the backaction force orientation"). We subsequently use Eq.~\ref{eq:M7b} for fitting the damping imbalance (solid line on Fig.~\ref{Fig3}(d), and further compute the expected centered frequency splitting given by Eq.~\ref{eq:M7a} to obtain the solid line shown on Fig.~\ref{Fig3}(e).

\bibliography{bib}
\end{document}


\title{\textsc{Supplementary Information}\\ Hyperspectral Electromechanical imaging at the nanoscale: Dynamical backaction, dissipation and quantum fluctuations}


\affiliation{CNRS, Inst. NEEL, "Nanophysique et semiconducteurs" group, 38000 Grenoble, France}
\affiliation{Université Paris-Saclay, CNRS, ENS Paris-Saclay, CentraleSupélec, LuMIn, 91405, Orsay, France}
\affiliation{Institut Universitaire de France, 1 rue Descartes, 75231 Paris, France}
\affiliation{Univ. Grenoble Alpes, CNRS, Grenoble INP, Institut N{\'e}el, F-38000 Grenoble, France}
\affiliation{Univ. Grenoble Alpes, CEA, IRIG, PHELIQS, “Nanophysique et semiconducteurs“ Group, F-38000 Grenoble, France}
\affiliation{Univ Lyon, ENS de Lyon, CNRS, Laboratoire de Physique, F-69342 Lyon, France}
\affiliation{School of Physics and Astronomy, University of Nottingham, Nottingham, NG7 2RD, United Kingdom}

\author{C. Chardin}
\affiliation{School of Physics and Astronomy, University of Nottingham, Nottingham, NG7 2RD, United Kingdom}

\author{S. Pairis}
\affiliation{Univ. Grenoble Alpes, CNRS, Grenoble INP, Institut N{\'e}el, F-38000 Grenoble, France}

\author{S. Douillet}
\affiliation{Univ. Grenoble Alpes, CNRS, Grenoble INP, Institut N{\'e}el, F-38000 Grenoble, France}

\author{M. Hocevar\,\orcidlink{0000-0001-5949-5480}} 
\affiliation{CNRS, Inst. NEEL, "Nanophysique et semiconducteurs" group, 38000 Grenoble, France}
\affiliation{Univ. Grenoble Alpes, CNRS, Grenoble INP, Institut N{\'e}el, F-38000 Grenoble, France}

\author{J. Claudon\,\orcidlink{0000-0003-3626-3630}}
\affiliation{Univ. Grenoble Alpes, CEA, INAC, PHELIQS, “Nanophysique et semiconducteurs“ Group, F-38000}

\author{J.-P. Poizat}
\affiliation{CNRS, Inst. NEEL, "Nanophysique et semiconducteurs" group, 38000 Grenoble, France}
\affiliation{Univ. Grenoble Alpes, CNRS, Grenoble INP, Institut N{\'e}el, F-38000 Grenoble, France}

\author{L. Bellon\,\orcidlink{0000-0002-2499-8106}}
\affiliation{Univ Lyon, ENS de Lyon, CNRS, Laboratoire de Physique, F-69342 Lyon, France}

\author{P. Verlot\,\orcidlink{0000-0002-5105-3319}}
\email{pierre.verlot@universite-paris-saclay.fr}
\affiliation{Université Paris-Saclay, CNRS, ENS Paris-Saclay, CentraleSupélec, LuMIn, 91405, Orsay,
France}
\affiliation{Institut Universitaire de France, 1 rue Descartes, 75231 Paris, France}

\date{\today}


\maketitle 

\section*{\underline{Supplementary Note}: Heat mediated electro-mechanical coupling}

\subsection{Heat equation}

The heat equation for the temperature field $T(y',t)$ along the nanowire is:
\begin{equation}
\partial_y' (\kappa \partial_y' T) = \rho c \partial_t T
\end{equation}
with $\kappa$ the thermal conductivity, $\rho$ the density and $c$ the heat capacity of InAs. The boundary conditions (BC) correspond to a fixed temperature $T_0$ at the origin, no heat flux at the nanowire free end, and a power influx $aP$ at position $(x,y)$:
\begin{subequations}
\begin{align}
T(0,t)&=T_0,\\
\kappa\partial_y' T(L,t)&=0,\\
\kappa\partial_y' T(y^-,t)-\kappa\partial_y' T(y^+,t)&=\frac{aP}{\pi R^2}.
\end{align}
\end{subequations}
Assuming $\kappa$ is constant, the static temperature profile $\Tbar$ solution of these equations is piecewise linear:
\begin{equation}
\Tbar(y')=T_0+\frac{\aPbar}{\pi R^2 \kappa}\min(y',y),
\end{equation}
where $\aPbar$ is the mean absorbed power from the e-beam. We now define $\Ttilde=T-\Tbar$ as the deviation from this stationary state, and study the dynamics of the temperature field due to non stationary part of the absorbed power $\aPtilde=aP-\aPbar$. In the Fourier space, the heat equation writes
\begin{equation} \label{EqDynHeat}
\partial_y'^2 \Ttilde = \alpha^2 \Ttilde, \ \text{with} \ \alpha^2=i\Omega \frac{\rho c}{\kappa},
\end{equation}
with the dynamic BC:
\begin{subequations}
\begin{align}
\Ttilde(0,\Omega)&=0,\\
\partial_y' \Ttilde(L,\Omega)&=0,\\
\partial_y' \Ttilde(y^-,\Omega)-\partial_y' \Ttilde(y^+,\Omega)&=\frac{\aPtilde(\Omega)}{\pi R^2 \kappa}.
\end{align}
\end{subequations}
We solve this equation with the ansatz $\Ttilde = A \sinh\alpha y'$ for $y'<y$ and $\Ttilde = B \cosh\alpha (y'-L)$ for $y'>y$ (satisfying the heat equation and the initial and final BC). The continuity of the temperature field in $y'=y$ imposes:
\begin{equation}
A \sinh \alpha y = B \cosh \alpha (y-L),
\end{equation}
while the discontinuity of the heat flux writes:
\begin{equation}
A \alpha \cosh \alpha y - B \alpha \sinh \alpha (y-L)= \frac{\aPtilde}{\pi R^2 \kappa}.
\end{equation}
We solve these two equations for $A$ and $B$ and conclude that the dynamic temperature field is:
\begin{align}
\Ttilde(x,y,y',\Omega) &= \thetatilde(x,y,\Omega) h\left(y,y',\sqrt{i \Omega\rho c/\kappa}\right),\\[1mm]
\thetatilde(x,y,\Omega) &= \frac{L}{\pi R^2 \kappa}\aPtilde(x,y,\Omega) ,\\[1mm]
\begin{split}
h(y,y',\alpha)&= H(y-y') \frac{\sinh \alpha y'  \cosh \alpha (y-L)}{\alpha L \cosh \alpha L}\\
& \quad +H(y'-y)\frac{\sinh \alpha y  \cosh \alpha (y'-L)}{\alpha L \cosh \alpha L}.
\end{split}
\end{align}
The dynamic temperature field is expressed here as the product of 2 terms, $\thetatilde(x,y,\Omega)$ with quantifies the amplitude of the forced temperature variation, and $h(y,y',\alpha)$ (where $\alpha$ is the complex number proportional to $\sqrt{i\Omega}$ defined in Eq.~\ref{EqDynHeat}) which accounts for the propagation of the damped thermal wave according to the heat equation.

\subsection{Temperature mediated force backaction mechanism}

When the temperature of the nanowire is changing, it can induce stresses that result in a change of the equilibrium local curvature $C=\partial_y'^2 \xi$, with $\xi(y')$ the lateral deflection. For example, let us assume the local curvature at temperature $T_0$ is $C_0$. When the temperature changes to $T$, the equilibrium curvature changes to $C=C_0/[1+\epsilon(T-T_0)]\sim C_0[1-\epsilon(T-T_0)]$, with $\epsilon$ the linear thermal expansion coefficient. More generally, we assume the effect to be small and define a linear thermal bending coefficient $\beta$, so that in equilibrium $C = C_0+ \beta (T-T_0)$. The equation of motion of the nanowire is a variation of Euler-Bernoulli's:
\begin{align}
\partial_y'^2 [EI(\partial_y'^2 \xi - \beta \Ttilde)] + \rho \pi R^2 \partial_t^2 \xi & = 0,\\
\partial_y'^2 [EI\partial_y'^2 \xi]+ \rho \pi R^2 \partial_t^2 \xi & = \partial_y'^2 (EI \beta \Ttilde), \label{EqEBcoupledT}
\end{align}
where we removed the static terms, and $EI$ is the flexural rigidity of the nanowire. The temperature field can thus be seen as an external force field
\begin{align}
f_\ba (x,y,y',\Omega) &= \partial_y'^2 [EI \beta(y') \Ttilde(x,y,y',\Omega)],\\
& = EI \thetatilde(x,y,\Omega) \partial_y'^2 [\beta(y') f(y,y',\alpha)].
\end{align}
We suppose that its effect on the mechanical response of the nanowire is small, so that the resonant mode shape $\phi(y')$ is unaffected by $f_\ba$. We can then project Eq.~\ref{EqEBcoupledT} on $\phi(y')$, solution of the unperturbed Euler-Bernoulli equation, and derive in the Fourier space the equation of the driven simple harmonic oscillator corresponding to the resonant mode:
\begin{equation}
(k - m \Omega^2 +i\gamma \Omega) \xi_L = F_\ba(x,y,\Omega).
\end{equation}

Here $m$ and $k$ are the lumped mass and stiffness of the oscillator, $\gamma$ has been introduced to describe the damping (unaccounted for in Eq.~\ref{EqEBcoupledT}), and the temperature mediated force backaction is:
\begin{align}
F_\ba(x,y,\Omega) &=\int_0^L f_\ba(x,y,y',\Omega) \phi(y') \dd y', \\
&=  \aPtilde(x,y,\Omega) \frac{EI L}{\pi R^2 \kappa} g(y,\Omega), \label{EqFba}
\end{align}
with
\begin{equation}
g(y,\Omega) = \int_0^L\partial_y'^2 [\beta(y') h(y,y',\sqrt{i \Omega \rho c /\kappa})] \phi(y') \dd y'. \label{Eqg}
\end{equation}

There are two origins for the variations of $\aPtilde$ that drive the dynamic temperature field. One is that the e-beam power fluctuates in time, thus $\Ptilde(\Omega)\neq 0$ and $\aPtilde=\abar\Ptilde$. The second is that the nanowire is deflected laterally with respect to the e-beam, changing the absorption $a$:
\begin{equation} \label{Eqatilde}
\atilde(x,y,\Omega)=\frac{\dd a}{\dd x}(x)  \xi(y,\Omega) = \frac{\dd a}{\dd x}(x) \frac{\phi(y)}{\phi(L)} \xi_L(\Omega),
\end{equation}
where the deflection amplitude $\xi(y,\Omega)$ at the position $y'=y$ of the e-beam is that of the nanowire free end $\xi_L$ scaled by the mode shape $\phi(y')$. In the following sections, we treat independently each of these two mechanisms.

\subsection{Backaction effects on the oscillator mechanical response}

In this section, we focus on the driving of the temperature field induced by the vibration of the nanowire: $\aPtilde=\atilde \Pbar \propto \xi_L(\Omega)$. This temperature field then generate bending in the nanowire, which in turn drive the vibration of the nanowire. The fields deformation-temperature are thus coupled, and the induced feedback modifies the mechanical responses of the system due to the e-beam measurement. This coupling can give rise to self oscillation or backaction cooling of the oscillator\cite{metzger2004cavity,ramos2012optomechanics,Nigues2014a,tsioutsios2017real}.

Injecting the expression of $\atilde$ from Eq.~\ref{Eqatilde} into Eq.~\ref{EqFba}, we can define a backaction stiffness $k_\ba$ with
\begin{align}
F_\ba(x,y,\Omega) &= k_\ba (x,y,\Omega) \xi_L(\Omega),\\
k_\ba (x,y,\Omega) & = \frac{\dd a}{\dd x}(x)  \Pbar \frac{EI L }{\pi R^2 \kappa} \frac{\phi(y)}{\phi(L)} g(y,\Omega). \label{Eqkba}
\end{align}
This electro-mechanical stiffness is a complex number, with a real part changing the effective stiffness of the oscillator, thus the resonance frequency, and an imaginary one modifying the damping. One noticeable feature is that $k_\ba$ is proportional to $\dd a / \dd x$, which is antisymmetric with respect to the lateral position $x$. The frequency shift and the modification in damping due to the backaction should be opposite on the left and right side of the nanowire.

To go further, we need to make some hypothesis on the spatial distribution of $\beta$. For example, a nanowire with a uniform natural curvature $C_0$, the linear thermal bending coefficient would be uniform: $\beta = -C_0 \epsilon$. The same uniformity would also apply to a nanowire coated on one side only and thus subject to the ``bimetal effect'' (differential expansion of its layers). In such case, we can extract $\beta$ from the integral in Eq.~\ref{Eqg} and end up with:
\begin{equation}
g(y,\Omega) = i \beta \Omega \frac{ \rho c}{\kappa}\int_0^L  h(y,y',\sqrt{i \Omega \rho c /\kappa}) \phi(y')\dd y'. \label{EqgBetaUniform}
\end{equation}
The dependency on the longitudinal coordinate $y$ of the backaction force can be estimated from this formula since the functions $h$ and $\phi$ are known.

Another hypothesis on $\beta$ is than it corresponds to one (or more) punctual defect(s): $\beta(y',\Omega)=\beta_d e^{i\Omega \tau_d} \delta (y'-y_d)$, with $y_d$ the position of the defect, $\beta_d$ the amplitude of the punctual thermal bending coefficient, and $\tau_d$ a microscopic response time of the mechanical stress to the temperature change. After a double integration by parts, Eq.~\ref{Eqg} leads to:
\begin{equation}
g(y,\Omega) = \beta_d e^{i\Omega \tau_d} h(y,y_d,\sqrt{i \Omega \rho c /\kappa}) \phi''(y_d). \label{EqgBetaPeaked}
\end{equation}
Here again, the spatial dependency of the backaction effect on $y$ can be predicted and compared to the experimental data. If several defects are required, all is required is to sum the functions $g$.

The effect of the distribution of $\beta$ on the backaction stiffness $k_\ba$ spatial profile is illustrated in Fig. S~\ref{Figk_ba}. The different hypotheses yield a very different signature for both the real and imaginary part. Experimental observations on the resonance frequency shift and the damping variation, both along and across the nanowire, should point at the best model to describe a specific sample.

\begin{figure}[htbp]
\begin{center}
\includegraphics{k_ba(y).pdf}
\caption{Backaction stiffness spatial profile for different hypotheses. The top panel presents the real part of the stiffness (modifying the resonance frequency), while the bottom panel presents the imaginary part (modifying the damping). The curves are normalized to maximum of $|k_\ba(y)|$ on all positions. The continuous curve corresponds to a uniform $\beta$ (Eq.~\ref{EqgBetaUniform}), the dashed and dotted ones to a single defect in $y_d/L=0.1$ or $y_d/L=0.5$ (Eq.~\ref{EqgBetaPeaked}, assuming $\tau_d=0$). The values of the parameters ($\Omega\rho c L^2/\kappa$) correspond to the nanowire of interest in the article.}
\label{Figk_ba}
\end{center}
\end{figure}

\subsection{Quantum backaction from power fluctuations}

In this section, we focus on the variations of $\aPtilde$ are due to the power fluctuations: $\aPtilde = \abar \Ptilde$. Eq.~\ref{EqFba} simply leads to:
\begin{align}
F_\ba(x,y,\Omega) &=  \abar (x) \Ptilde(\Omega) \frac{EI L}{\pi R^2 \kappa} g(y,\Omega), \label{EqFbaQ}
\end{align}
Since $g(y,\Omega)$ is the same function as in the previous section, the same hypotheses (uniform distribution or punctual defect) will lead to the same conclusions for the spatial dependency $g(y,\Omega)$. Fig. S~\ref{Fig_g_y_Omega} reports the shape of the force profile expected in the different cases. Once again, the behavior is very different and should be helpful to discriminate among the models for specific samples.

\begin{figure}[htbp]
\begin{center}
\includegraphics[width=\columnwidth]{g(y,Omega).pdf}
\caption{Quantum backaction force spatial profile for different hypotheses. The continuous curve corresponds to a uniform $\beta$ (Eq. \ref{EqgBetaUniform}), the dashed and dotted ones to a single defect in $y_d/L=0.1$ and $y_d/L=0.5$ (Eq. \ref{EqgBetaPeaked}, assuming $\tau_d = 0$).
The values of the parameters $\Omega\rho cL^2/\kappa$ correspond to the nanowire of interest in the article.}
\label{Fig_g_y_Omega}
\end{center}
\end{figure}

\section*{\underline{Supplementary Note}: Determination of the backaction force orientation}

This section explains how the orientation of the backaction force is determined. The principle relies on extracting the dissipative component of their gradient from measuring the effective mechanical damping rates, which are essentially insensitive to the permanent, measurement-induced changes experienced by the NW:

\begin{eqnarray}
\Gamma_{\mathrm{eff},1}(y)&=&\Gamma_0+\frac{\mathrm{Im}\left(\partial_x F_{\mathrm{ba}}\right)}{m\Omega_0}\cos{\theta_{\mathrm{ba}}}\label{eq:gamma1}\\
\Gamma_{\mathrm{eff},2}(y)&=&\Gamma_0+\frac{\mathrm{Im}\left(\partial_x F_{\mathrm{ba}}\right)}{m\Omega_0}\sin{\theta_{\mathrm{ba}}}.\label{eq:gamma2}
\end{eqnarray}

The angle $\theta_{\mathrm{ba}}$ between the first vibrational direction and the backaction force is subsequently given by the ratio of the effective damping deviations with respect to the intrinsic damping value, $\theta_{\mathrm{ba}}=\arctan\left(\frac{\Gamma_{\mathrm{eff},2}-\Gamma_0}{\Gamma_{\mathrm{eff},1}-\Gamma_0}\right)$. Figure S\ref{Fig_rawdamp} shows the longitudinal evolution of the damping deviations associated with modes 1 and 2 (red and blue curves, respectively) across the transverse position $x/d=0.46$. The solid lines are theoretical fits given by Eqs. \ref{eq:gamma1} and \ref{eq:gamma2}, with $F_{\mathrm{ba}}$ being set by Eq. 4 from the main manuscript. The fluctuations of the data likely result from a residual sensitivity of the intrinsic damping rate with respect to the e-beam measurement process, which however represents a common mode of noise and cancels when taking the damping imbalance presented on Fig. 3(d) of the main manuscript. The ratio of the amplitudes of these two curves yields $\theta_{\mathrm{ba}}\simeq-\SI{0.6}{rad}$.

\begin{figure}[htbp]
\begin{center}
\includegraphics[width=\columnwidth]{damping-raw.pdf}
\caption{Longitudinal evolution of the respective damping rates deviations, for mode 1 (blue) and mode 2 (red), respectively. The solid lines correspond to the model given by Eq. \ref{eq:gamma1} and \ref{eq:gamma2}, respectively.}
\label{Fig_rawdamp}
\end{center}
\end{figure}

\bibliography{bib}